\documentclass[12pt,preprint]{aastex}

\def\lsim{\hbox{ \rlap{\raise 0.425ex\hbox{$<$}}\lower 0.65ex\hbox{$\sim$} }}
\def\gsim{\hbox{ \rlap{\raise 0.425ex\hbox{$>$}}\lower 0.65ex\hbox{$\sim$} }}

\def\opeqn{\begin{equation}}
\def\cleqn{\end{equation}}

\begin{document}

\title{ Addendum to Astro-ph/0207612: Reflection Symmetry of Cusps in 
Gravitational Lensing }
\author{Sun Hong Rhie (UND)}

%\centerline{(\today)}

\begin{abstract}
A contour plot of positive iso-$J$ curves is shown for a gravitational
binary  lens $\epsilon_2 = 0.1883$ and $\ell = 0.687$. 
The caustic curve is made of 4-cusped central caustic (so-called 
``stealth bomber") on the lens axis and two triangular caustics off 
the lens axis. The cusps of the trioids are all negative cusps. The positive
image magnification contours outside the positive cusps on the lens axis
are elongated along the symmetry axis.  It is in contrast with figure 4
of astro-ph/0206162 (Gaudi and Petters) where the contours appear to contract 
along the symmetry axis. We apologize if our citation of their figure 4 without
the caution played any role of causing confusion with readers.  
\end{abstract}


\clearpage

\begin{figure}
\plotone{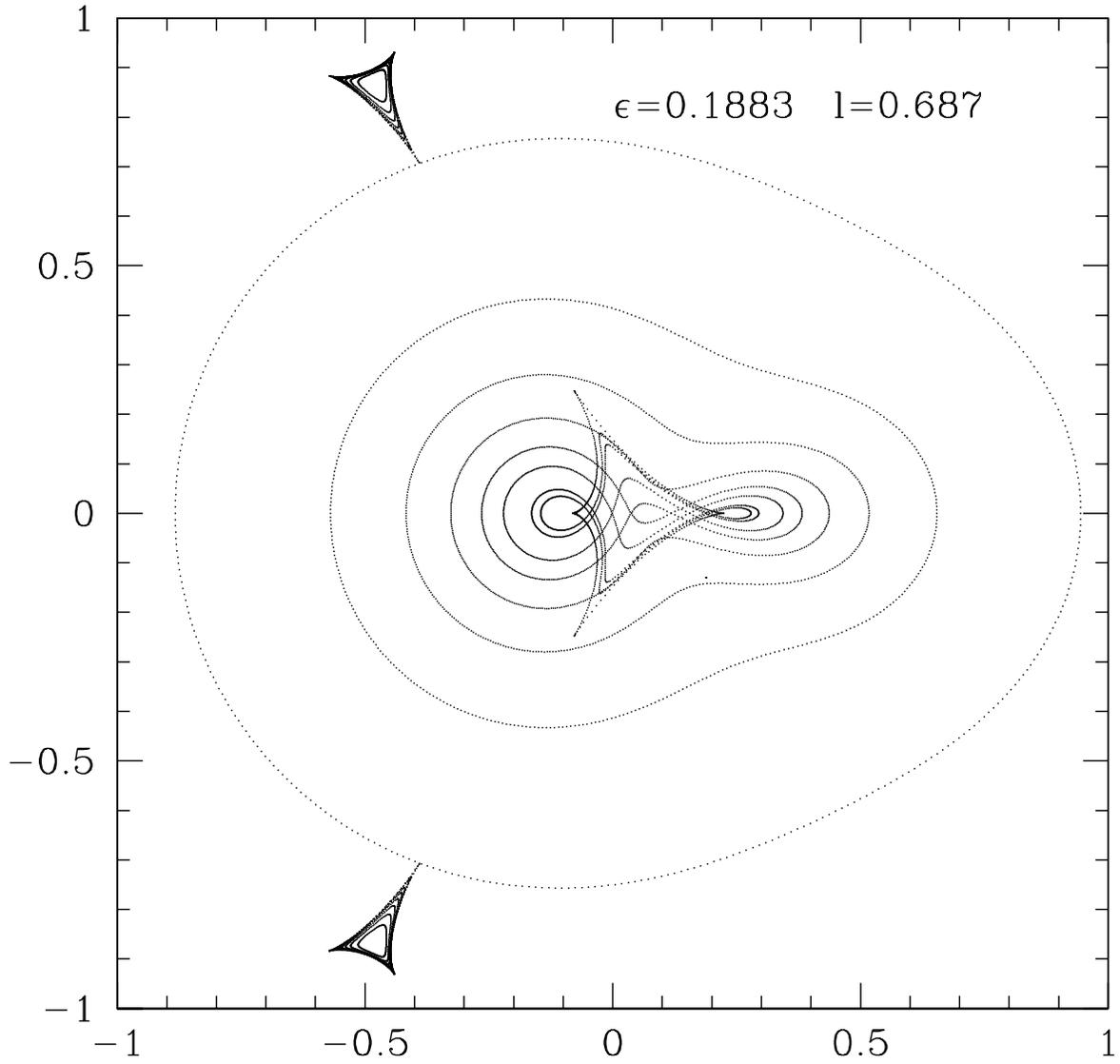}
\figcaption{\label{fig-add} 
Iso-$J$ curves of a binary lens $\epsilon=0.1883$ and $\ell=0.687$ where
$J = 0$ and $J = 10^{0.1 n -1}: ~n = 0, 1, 2, ... , 8$.  
}
\end{figure}

In microlensing experiments, the light curves are the primary sources of 
information, and the singular behavior of the caustics commands attention.  
It is rather amusing that the weired shapes of the caustics with spiky 
cusps faithfully manifest themselves as towering light curves. We may say
a pleasure of studying reality through precision illusion. If we imagine
a line in the plane in figure \ref{fig-add}, the magnification pattern along 
the line consitutes a light curve, with a caveat that the motion of the observer
(such as on the earth) or the orbital motion of the binary objects can cause 
variations. In astro-ph/0207612, we showed an example of a connected caustic
and also briefly discussed diamond caustics which are typical of binary lenses 
with relatively large separations. We discussed that cusps off the lens axis of 
a connected caustic are negative cusps. Here we show that that is true for
all binary lenses. Namely, the trianguar caustics have cusps off the lens
axis and the cusps are all negative cusps.  The positive iso-$J$ contours line 
the triangular caustics from inside. As $J$ converges to the maximum value 1, 
the families of iso-$J$ contours shrink to the points that generate images 
at the finite limit points where $J =1$. Near the cusps, the lens equation 
becomes cubic because there are three images with small values of the Jacobian 
determinant $J$, two of which can be unrealized images (or complex solutions
to the cubic equation). 

When a source trajectory approaches a caustic, the 
light curve can be more dominated by the behavior of the line caustic or more
by a cusp. The distinction is made by the singlet image outside the 
non-critical precaustic curve (near a positive cusp, inside if a negative cusp),
and the magnification contours look like ``balloons" generally elongated    
along the symmetry axis. Figure 4 in astro-ph/0206162 by Gaudi and Petters 
(Fig-GP) we cited in our paper astro-ph/0207612 shows a different pattern.  
Since the plot Fig-GP is drawn without numerical specifications, we only can 
see the general pattern and the outstanding difference seems to be that    
the contours contract along the symmetry axis of the cusp. 
We suspect that most of the contours of Fig-GP are out of the validity range of 
the cubic equation which the lens equation is reduced to near a cusp. 
For example, the positive singlet image can not have magnification less 1
because the Jacobian determinant is globally bounded by $J \le 1$ and the
magnification is $1/J$. The figure caption indicates that the magnification 
range of the iso-$J^{-1}$ curves is $0.1 < |J|^{-1} < 10$. Since the adjacent 
contours differ by the ratio $J_1/J_2 = 10^{0.1}$, 10 large outer contours of 
the positive image can not exist. Thus, the contracting contour pattern of 
Fig-GP along the symmetry axis may be best understood as an illustration of 
the limitation of the cubic equation away from the cusp. (Gaudi assured us 
that the plot was drawn properly from equation 9 of astro-ph/0206162.)   
Figure \ref{fig-add} shows an iso-$J$ contour plot of the same binary lens   
$\epsilon=0.1883$ and $\ell=0.687$ whose parameters Gaudi provided to us.

\clearpage

%\onecolumn

% figures follow here
%
% Here is an example of the general form of a figure:
% Fill in the caption in the braces of the \caption{} command. Put the label
% that you will use with \ref{} command in the braces of the \label{} command.
%

\end{document}